\documentclass{ws-ijbc}
\newcommand{\wsdraftnote}{}
\usepackage{ws-rotating}     % used only when sideways tables/figures are used
\usepackage{xcolor}
\usepackage[verbose]{hyperref}
\usepackage{natbib} 
\usepackage{verbatim}
\usepackage{subcaption}
\hypersetup{colorlinks=false,allbordercolors=blue,pdfborderstyle={/S/U/W 1}}
\begin{document}

% \catchline{}{}{}{}{} % Publisher's Area please ignore

%\markboth{Authors' Names}{Paper Title}

\title{How Network Topology Affects the Strength of Dangerous \\ Power Grid Perturbations}
%\footnote{For the title, try not to use more than
%three lines. Typeset the title in 15 pt Times Roman and boldface.}}

\author{Calvin Alvares}
%\footnote{Typeset names in 11 pt Times Roman.
%Use the footnote to indicate the present or permanent address of
%the author.}}

\address{Department of Physical Sciences,\\ Indian Institute of Science Education and Research Kolkata,\\ Mohanpur, 741 246, India\\
alvarescalvin16@gmail.com}

% \footnote{State completely without
% abbreviations the affiliation and mailing address, including
% country. Typeset in 11~pt Times Italic.}}

\author{Soumitro Banerjee}
\address{Department of Physical Sciences,\\ Indian Institute of Science Education and Research Kolkata,\\ Mohanpur, 741 246, India\\
soumitro@iiserkol.ac.in}

\maketitle

% \begin{history}
% \received{(Month Day Year)}; \accepted{(Month Day Year)}; \published{(Month Day Year)}
% \end{history}

\begin{abstract}
Reasonably large perturbations may push a power grid from its stable synchronous state into an undesirable state. Identifying vulnerabilities in power grids by studying power grid stability against such perturbations can aid in preventing future blackouts. Probabilistic stability quantifiers such as \textit{basin stability}, which measures the {asymptotic} stability {of a system, and \textit{survivability}, which measures the transient stability of a system}, have been commonly used to quantify the stability of nodes in a power grid. However, these quantifiers do not provide information about the strength of perturbations that destabilize the system. To measure the strength of perturbations beyond which the stability of the system gets compromised,  we employ two probabilistic distance-based stability measures --- \textit{basin stability bound}, which deals with a system's asymptotic behaviour, and \textit{survivability bound}, a newly defined stability measure that deals with a system's transient behaviour. Using these stability quantifiers, we conduct a detailed study on the impact of network topology on the strength of dangerous power grid perturbations. In this, we uncover a new class of highly vulnerable nodes that were previously unknown. Additionally, we establish connections with tree-like network structures and node connectivity to lowly stable nodes.

\end{abstract}


%\begin{multicols}{2}
\vspace*{-0.5in}

\section{Introduction}
Power grids are critical infrastructures that underpin the functioning of modern society. Failures in these systems have resulted in large-scale blackouts that have left millions of people without electricity \cite{ind_blackout, turkey_blackout, italy_blackout}. In recent times, expansion, modernization, and decentralization of the grid have been promoting rapid changes in existing power grid infrastructure \cite{pepermans2005distributed, di2018decarbonization, fang2011smart}. Thus, as power grids become increasingly complex, it is important to ensure that they are resilient to various perturbations to prevent future blackouts. 

During normal operation, all parts of a power grid function at the same frequency \cite{machowski2020power}. This state is called the grid's synchronous state. Perturbations such as a line being switched off or the power demand of the grid not being met may result in large parts of the grid desynchronizing, causing cascading failure \cite{buldyrev2010catastrophic, motter2002cascade, schafer2018dynamically}. As disturbances affecting power grids can be reasonably large, linear stability analysis by means of evaluating the eigenvalues of the Jacobian matrix at an equilibrium point or the master stability function \cite{pecora1998master} cannot be employed as a measure of grid stability against such perturbations. 

To quantify the stability of dynamical systems, such as power grids, against reasonably large perturbations, several non-local stability measures have been proposed \cite{menck2013basin, hellmann2016survivability, mitra2015integrative, mitra2017multiple, klinshov2015stability, halekotte2020minimal, klinshov2018interval}. Among these, a popular stability measure, known as {\em basin stability} \cite{menck2013basin}, relates to the fraction of phase space that forms the basin of attraction of the desirable attractor. In the context of power grids, the desirable attractor corresponds to the stable synchronous state of the grid. Basin stability has been used extensively in the study of power grid stability \cite{menck2014dead, ji2014basin, schultz2014detours, nitzbon2017deciphering, kim2016building, nauck2022predicting, feld2019large, schafer2016taming}.

Basin stability deals with the asymptotic, or, long-term behavior of a system. In addition to asymptotic behavior, transient behavior, that is, the behavior of a system before it reaches its steady state, is particularly relevant when dealing with power grids. Power grids operate within certain frequency bounds, and control mechanisms are triggered when a disturbance causes the grid to operate outside the set frequency bound. Such perturbations are undesirable for the system. A stability measure called {\em survivability} has been proposed to quantify the transient stability of dynamical systems \cite{hellmann2016survivability}. The set of states that do not leave a given desirable region of the phase space within a given time is called the {\em basin of survival}, and survivability is the fraction of states that are part of the system's basin of survival. Thus, survivability measures a system's {capacity} to {resist perturbations from amplifying past the system's desirble region. } Survivability, like basin stability, has been used to study the stability of the electricity grid \cite{hellmann2016survivability, nitzbon2017deciphering, buttner2022ambient}.

Basin stability and survivability are both volume-based measures of stability as they are related to the volume of the basin of attraction and the volume of the basin of survival, respectively. For power grids, the basin of attraction of the synchronous state can be very distorted \cite{halekotte2021transient}. It is possible that a stronger perturbation is safe for the system, but a weaker one is not. Hence, it is crucial to understand the magnitude of perturbations that are dangerous to the system. Volume-based stability measures fail to capture this. However, \textit{basin stability bound} \cite{alvares2024probabilistic}, a distance-based stability measure that provides a safe bound to the strength of perturbations based on the asymptotic behavior of the system, captures this aspect. Analogous to basin stability bound, we propose a distance-based stability quantifier called \textit{survivability bound} that provides a safe bound to the strength of perturbations based on the system's transient behavior. Although, in this paper, we use the above methods on power grids, they are readily applicable to other dynamical systems.

Using the two distance-based stability measures, we conducted a detailed study on the effect of network topology on the stability of the power grid. Nodes with a low basin stability bound indicate that a low perturbation strength is required to cause permanent loss of grid synchrony. On the other hand, nodes with a low survivability bound indicate that a low perturbation strength is required to cause undesirable transients in the system, which do not necessarily lead to loss of grid synchrony.  In this study, based on distance-based stability quantifiers and a novel method of quantifying the strength of power grid perturbations, we outline the local network topology that results in low single-node stability {and uncover a previously unknown, new class of highly vulnerable nodes.}

This paper is organized as follows. The power grid model, methods of stability quantification, and a classification scheme of network nodes are presented in section~\ref{sec:methods}. Section \ref{sec:results} describes the results, and section \ref{sec:conclusion} concludes the work. 

\section{Methods}\label{sec:methods}
\subsection{Power grid model}
We use a complex network representation to model power grids with generators and consumers as nodes and transmission lines as edges. Generators and consumers are modeled as synchronous machines that follow the swing equation \cite{machowski2020power}. The equations that describe the dynamics of the grid are \cite{filatrella2008analysis}
%\numparts
    \begin{eqnarray}
    \dot{\theta_i} &=& \omega_i, \label{eq:eqswing_a}\\
    I_i \dot{\omega_i} &=& -\gamma_i \omega_i + \bar{P_i} - \sum^N_{j=1} T_{ij}\sin(\theta_i - \theta_j),\label{eq:eqswing_b}
\end{eqnarray}
%\endnumparts
where $\theta_i$ and $\omega_i$ are the phase angle and the angular velocity of the synchronous machine at the $i^{\rm th}$ node of the power grid network in a frame rotating at the grid frequency. $I_i$ is the inertia at the $i^{\rm th}$ node, $\gamma_i$ is the damping at the $i^{\rm th}$ node, and $\bar{P}_i$ is proportional to the net power generated or consumed at the $i^{\rm th}$ node. $T_{ij}$ is the transmission capacity between node $i$ and node $j$. If node $i$ and node $j$ are not connected, then $T_{ij} = 0$. 

If all nodes have equal inertia ($I_i = I$) and damping ($\gamma_i = \gamma$), then the equations can be simplified as 
%\numparts 
    \begin{eqnarray}
    \dot{\theta_i} &=& \omega_i, \label{eq:eqswing_simple_a}\\
    \dot{\omega_i} &=& -\alpha \omega_i + P_i - \sum^N_{j=1} K_{ij}\sin(\theta_i - \theta_j),, \label{eq:eqswing_simple_b}
    \end{eqnarray}
%\endnumparts
where $\alpha = \gamma/I$, $P_i = \bar{P_i}/I$, $K_{ij}=T_{ij}/I$.

In this paper, we use the simplified differential equations described by equations (\ref{eq:eqswing_simple_a}) and (\ref{eq:eqswing_simple_b}) to model power grids. Each node in the network has two corresponding dynamical variables --- a phase angle and a frequency. The fixed point of these equations corresponds to the stable synchronous state of the grid. In this state, the $i^{\rm th}$ node has a phase $\theta_i^s$ and frequency $0$ in the rotating frame of reference.

\subsection{Basin stability bound}

We first describe the basin stability introduced in \cite{menck2013basin, menck2014dead}.

Consider an $N$ dimensional dynamical system with a phase space $X$. The set $\mathcal{A}$ is the system's desirable attractor. This attractor has a basin of attraction $\mathcal{B}$. {Basin stability for the attractor $\mathcal{A}$ is defined in a finite region of the phase space.}

The basin stability of the attractor $\mathcal{A}$, defined in {this finite} region $X^P$ is the fraction of states in the region $X^P$ contained in the attractor's basin of attraction $\mathcal{B}$. The basin stability, assuming a uniform distribution of perturbations, is defined as 
\begin{equation}
    \beta (X^P) = \frac{\textnormal{Vol}( \mathcal{B} \cap X^P)}{\textnormal{Vol}(X^P)}
\end{equation}
Thus, basin stability measures the probability that a perturbation in the region $X^P$ causes the system to return to its desirable attracting state. 

To define basin stability bound \cite{alvares2024probabilistic}, we consider a finite subset of the phase space $X^0 \subseteq$ X, representing the extent to which perturbations can push the system. {Basin stability bound is defined in this region $X^0$.}

Let ${X^D(d)}$ be the set of states within a distance $d$ from the attractor $\mathcal{A}$ that lie in the set $X^0$, i.e., 
\begin{equation}\label{eq:distregion}
    {X^D}(d) = \{x \in {X^0} \:|\: \textnormal{dist} (x,\mathcal{A})<d\}
\end{equation}
where $\textnormal{dist}(x,\mathcal{A})$ is the distance of the state $x$ to the attractor $\mathcal{A}$. {In the case of the power grid model used in this paper, the attractor $\mathcal{A}$ is a point attractor. The distance function $\textnormal{dist} (x,\mathcal{A})$ used in this paper is defined in section \ref{sec:dist}.}

\noindent ${D_{\beta}}$ is the set of distances at which the corresponding basin stability is less than a basin stability tolerance $t_\beta$. Thus,
\begin{equation}\label{eq:setdist}
    {D_{\beta}} = \{d \in (0, d_\textnormal{max}] \:|\: \beta ( {X^D}(d)) < t_\beta\}
\end{equation}
where $t_\beta \in (0,1]$ is a predefined basin stability tolerance, and $d_{\textnormal{max}}$ is the maximum distance we would like to consider.

\noindent The basin stability bound of the attractor $\mathcal{A}$ is defined as \cite{alvares2024probabilistic}
\begin{equation}\label{eq:stabbound}
    \bar{\beta} =
    \begin{cases} 
        \inf(D_\beta) & \text{if } D_\beta \neq \varnothing \\
        d_{\text{max}} & \text{otherwise}
    \end{cases}
\end{equation}

Thus, the basin stability bound is the minimum distance at which the corresponding basin stability is less than the tolerance $t_\beta$.

Single-node basin stability bound is defined as the basin stability bound with perturbations applied at a single network node starting from the initial attracting state. For a power grid, we consider a perturbation to $\theta_i$ as a single-node perturbation at node $i$. If single-node perturbations occur in the region $X^{sn}$, such that $\theta_i \in X^{sn}$ is a single-node perturbation at the $i^{\rm th}$ node; then the single-node basin stability bound of node $i$ of the grid network is the basin stability bound defined in the region.
\begin{eqnarray}
\label{eq:singnoderegion}
X_i^0 = \{ (\theta, \omega) \in X | (\theta_i \in X^{sn} \land \omega_i = 0 \land \\ (\forall j \neq i: \theta_j = \theta^s_j \land \omega_j = 0) \}
\end{eqnarray}
where $\theta = (\theta_1, \theta_2, \cdots \theta_N)$ and $\omega = (\omega_1, \omega_2, \cdots \omega_N)$

Basin stability is estimated using a Monte Carlo simulation. In the region ${X^P}$, a number of initial conditions, $n$, are sampled from a uniform distribution. If the number of initial conditions that converge to the attractor $\mathcal{A}$ is $n_{\beta}$, then the estimated basin stability of the attractor is 
\begin{equation}
    \hat{\beta} (X^P) = \frac{n_{\beta}}{n}
\end{equation}l

To compute the basin stability bound, the basin stability $\hat{\beta}(X^D (d))$ is computed from $d = d_\textnormal{max}$ to $d = d_0$, using $n$ samples for every basin stability estimation. $d_0$ is the largest distance such that $\hat{\beta}({X^D} (d_0))=1$. For $\hat{\beta} (X^D(d))<t_\beta$, the values of $d$ are noted and added to a set ${D_\beta}$. The basin stability bound is computed using equation \ref{eq:stabbound}. Refer to the paper by Alvares et al. \cite{alvares2024probabilistic} for a detailed computation procedure and the error associated in basin stability bound's estimation.

\subsection{Survivability bound}

Suppose that the system has a desirable region $X^{+} \subseteq X$ of the phase space. A perturbation is considered safe if the state does not leave the desirable region in finite time $t$. Let $X^S_t$ be the set of points in $X^P$ that do not leave the desirable region $X^+$ in the time $t$. Assuming a uniform distribution of perturbations in the region $X^P$, survivability is defined as \cite{hellmann2016survivability, nitzbon2017deciphering}
\begin{equation}
    \sigma (X^P) = \frac{\textnormal{Vol}(X_t^S)}{\textnormal{Vol}(X^0)}
\end{equation}
Survivability, thus, represents the probability that a perturbation in the region $X^P$ remains in the desirable region.

Consider ${D_{\sigma}}$ the set of distances at which the corresponding survivability is less than a survivability tolerance $t_\sigma$.
\begin{equation}
    {D_{\sigma}} = \{d \in (0, d_\textnormal{max}] \:|\: \sigma ( {X^D}(d)) < t_{\sigma}\}
\end{equation}
where $t_{\sigma} \in (0,1]$ is a predefined survivability tolerance, $d_{\textnormal{max}}$ is the maximum distance we would like to consider, and $X^D(d)$ is given by equation (\ref{eq:distregion}).

\noindent We define the survivability bound of the attractor ${\mathcal{A}}$ as
\begin{equation}\label{eq:eq5}
    \bar{\sigma} =
    \begin{cases} 
        \inf(D_\sigma) & \text{if } D_\sigma \neq \varnothing \\
        d_{\text{max}} & \text{otherwise}
    \end{cases}
\end{equation}

Thus, the newly proposed stability quantifier ---survivability bound is the minimum distance at which the corresponding survivability is less than the tolerance $t_\sigma$. 

In the case of power grids, the desirable region is defined to be 
\begin{equation}
X^+ = \{ (\theta, \omega) \in X |  \forall i: -\pi<\theta_i<\pi \land -\omega^+<\omega_i < \omega^+ \}
\end{equation}
where $\theta = (\theta_1, \theta_2,..., \theta_N)$ and $\omega = (\omega_1, \omega_2,..., \omega_N)$ and $w^+$ is a set bound to the frequency fluctuation. 

If single-node perturbations occur in the region $\theta_i \in X^{sn}$, then the single-node survivability bound of node $i$ is the survivability bound defined in the phase space region given by equation (\ref{eq:singnoderegion}).

Survivability is estimated using a Monte Carlo simulation. In the region ${X^P}$, a number of initial conditions, $n$, are uniformly sampled. If the number of initial conditions that converge to attractor $\mathcal{A}$ is $n_{\sigma}$, then the estimated survivability is
\begin{equation}
    \hat{\sigma} = \frac{n_{\sigma}}{n}
\end{equation}
The survivability bound is computed using the same procedure as used to compute the basin stability bound.

\subsection{Quantifying the strength of power grid perturbations} \label{sec:dist}
Basin stability bound and the survivability bound rely on the notion of distance between a perturbed state and the attractor. This distance is indicative of the strength of the perturbation. The distance in the phase space of a dynamical system can be quantified in various ways, ranging from standard Euclidean distance to the energy difference between the two states \cite{alvares2024probabilistic, lundstrom2018find}. For the distance of a perturbation from the synchronous state to have a physical meaning, we quantify it by defining an energy function that is related to the energy change of the system due to the perturbation.

Note that we have used a simplified power grid model given by equations (\ref{eq:eqswing_simple_a}) and (\ref{eq:eqswing_simple_b}) with the inertia at all nodes being the same and quantities such as torque and energy that we refer to are normalized by inertia. 

Consider a perturbation in the phase angle $\theta_i$ at the node $i$ from the stable state value of $\theta_i = \theta_i^s$ to $\theta_i = \theta_i^p$. It can be that either $\theta_i^s<\theta_i^p$ or $\theta_i^s>\theta_i^p$. We assume that no energy is transferred back from the system in the process of perturbing $\theta_i$. The work $W(\theta_i^s, \theta_i^p)$ done in moving from $\theta_i^s$ to $\theta_i^p$ is
\begin{equation}
    W(\theta_i^s, \theta_i^p) = \int^{\theta_i^p}_{\theta_i^s} \tau(\theta_i)\:d\theta_i 
\end{equation}
where, as per the above assumption,
\begin{equation}
    \tau (\theta_i) = 
    \begin{cases}
        -\textnormal{min}\{0,  \frac{d^2\theta_i}{dt^2}\} & \textnormal{if } \theta_i^p>\theta_i>\theta_i^s \\- \text{max}\{0, \frac{d^2\theta_i}{dt^2}\} &\text{if } \theta_i^p<\theta_i<\theta_i^s
    \end{cases}
\end{equation}
Additionally, we also assume that the perturbation is made at the synchronous grid frequency, which makes $w_i=0$. With this assumption,
\begin{equation}
    \frac{d^2\theta_i}{dt^2} = P_i - \sum^N_{j=1} K_{ij}\sin(\theta_i - \theta_j)
\end{equation}

The perturbation from $\theta_i^s$ to $\theta_i^p$ can happen through two different paths. For $\theta_i^p > \theta_i^s$, the two possible paths are $\theta_i^s \rightarrow \theta_i^p$ and $\theta_i^s \rightarrow -2 \pi+\theta_i^p $. For $\theta_i^p < \theta_i^s$, the two possible paths are $\theta_i^s \rightarrow \theta_i^p$ or $\theta_i^s \rightarrow 2 \pi+\theta_i^p $. We define energy $E(\theta_i^p)$ as the energy of the perturbation from $\theta_i^s$ to $\theta_i^p$ along the path corresponding to the least energy change. Thus,
\begin{equation}
    E(\theta_i^p) = 
    \begin{cases}
         \textnormal{min}\{W(\theta_i^s, \theta_i^p), \: W(\theta_i^s, -2\pi + \theta_i^p) \} & \text{if }  \theta_i^p > \theta_i^s \\
         \text{min}\{W(\theta_i^s, \theta_i^p), \: W(\theta_i^s, 2\pi + \theta_i^p) \} & \text{if }  \theta_i^p < \theta_i^s
    \end{cases} 
\end{equation}

The distance between the perturbed state and the attractor is taken as the energy change of the system corresponding to a perturbation in the phase $\theta_i^p$ at node $i$ from the initial stable state. Thus, 
\begin{equation}
\textnormal{dist}(x_i^p, \mathcal{A}) =   E(\theta_i^p) 
\end{equation}
where $x_i^p = (\theta_1^s,\: \theta_2^s,\: ...,\: \theta_i^p,\: ..., \:\theta_N^s,\: 0, \:0, \:..., \:0,\: ...,\: 0)$ 

\subsection{Network motifs}\label{sec:mot}
J. Nitzbon et al. \cite{nitzbon2017deciphering} have identified some important network motifs relevant to studying power grid stability. These network motifs are described below.

Consider a graph $G=(V,E)$ with vertices $V$ and edges $E$. $G=(V,E)$ is a tree if it is connected and has no cycles. $T'=(V',E')$ is a tree-shaped part of $G$ if it is an induced subgraph of $G$, is a tree and is maximal with the property that there is exactly one node $r \in V'$ that has at least one neighbor in $G-T'$. The node $r$ of $T'$ is called a root node ($R$). The union of all tree-shaped parts of the graph $G$ is called the forest $T$ of the graph $G$. For a tree-shaped part, the depth of node $x$ is the length of the shortest path from node $x$ to the root node. 

\begin{figure}[tbh]
    \centering
    \includegraphics[scale=0.6]{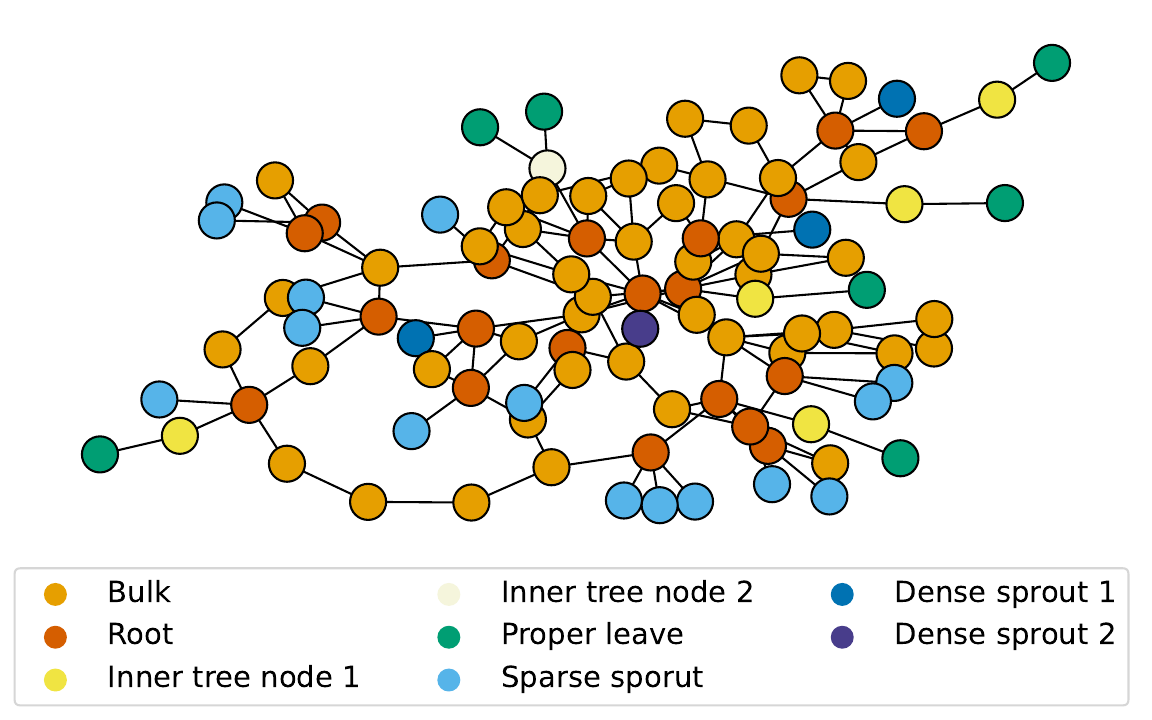}
    \caption{A network with nodes classified as bulk nodes, root nodes, inner tree nodes (class 1 and class 2), proper leaves, sparse sprouts, and dense sprouts (class 1 and class 2).}
    \label{fig:typesofnodes}
\end{figure}

Nodes that do not belong to the forest part $T$ are called bulk nodes ($B$). Nodes that belong to the forest part and are not root nodes are called non-root nodes. Non-root nodes that have a degree greater than one are called inner tree nodes ($I$). Non-root nodes that have a degree of one are called leaves. Leaves that have a depth of more than one are proper leaves ($P$), and leaves that have a depth of one are called sprouts. Sprouts with an average neighbor degree of less than six are called sparse sprouts ($S_s$), and sprouts with an average neighbour degree of more than five are called dense sprouts ($S_d$). Fig.~\ref{fig:typesofnodes} shows a network with nodes classified as described above.

In addition to the above classification scheme, we further divide the inner leave nodes into two classes. Class 1 inner tree nodes are inner tree nodes with a degree of two, and class 2 inner leaf nodes are inner tree nodes with a degree greater than two. We also divide dense sprouts into two classes. Class 1 dense sprouts are dense sprouts with a degree of six and seven, and class 2 dense sprouts have a degree greater than six.

\section{Results}\label{sec:results}

We use a random network generator model proposed by Schultz et al. \cite{schultz2014random} to generate realistic power grid networks to study the effect of network topology on power grid stability. The parameters chosen for this model are $N_0 = 1$, $p=1/5$, $q=3/10$, $s=1/3$, $t=1/10$. $50$ such networks, each consisting of $100$ nodes, were generated. In each network, half of the nodes were taken to be generators, and half of the nodes were taken to be consumers. Grid networks were modelled using equations (\ref{eq:eqswing_simple_a}) and (\ref{eq:eqswing_simple_b}), with the following parameters: $K=6$ for every transmission line, $\alpha = 0.1$ for every node, $P=1$ for every generator, and $P=-1$ for every consumer. One unit of time in the differential equations corresponds to $0.25$ s. Using this ensemble of grid networks, we investigate the single-node basin stability bound and the single-node survivability bound for every node of all the networks. 

Deviations in the grid frequency of power grids are generally kept within $\pm1$ Hz. A deviation of $\pm1$ Hz in the frequency corresponds to a $\omega \approx \pm 1.57$ in the units we have used. We choose two values for the desirable region's frequency bound $\omega^+$. We choose $\omega^+=1$ corresponding to an allowed frequency deviation of $\pm 0.64$ Hz. Another value $\omega^+=5$, corresponding to an allowed frequency deviation of $\pm3.2$ Hz, is chosen to investigate how the survivability bound changes when large frequency deviations are allowed. The region $X^{sn} = [-\pi,\pi]$ represents single-node perturbations in the phase angle. The single-node basin stability bound and single-node survivability bound of node $i$ in a network are computed in the region $X_i^0$ given by equation (\ref{eq:singnoderegion}) for single-node perturbations in the region $X^{sn}$. Basin stability bound and survivability bound are computed using tolerances $t_{\beta} = 0.95$ and $t_{\sigma} = 0.95$, respectively, and the maximum perturbation distance limit is set to $d_{\rm max} = 50$. For all basin stability and survivability computations, $n = 200$ initial conditions were used, and every initial condition was evolved for time $t=100$. Numerically, a perturbation is counted to be in the basin of attraction if $|\omega_i| < 1 \:\forall i$ at $t=100$. 

\begin{figure}[tbh]
    \centering
    \begin{subfigure}{0.32\textwidth}
    \centering
    \caption{}
    \label{fig:stab_deg}
    \includegraphics[width = 0.95\textwidth]{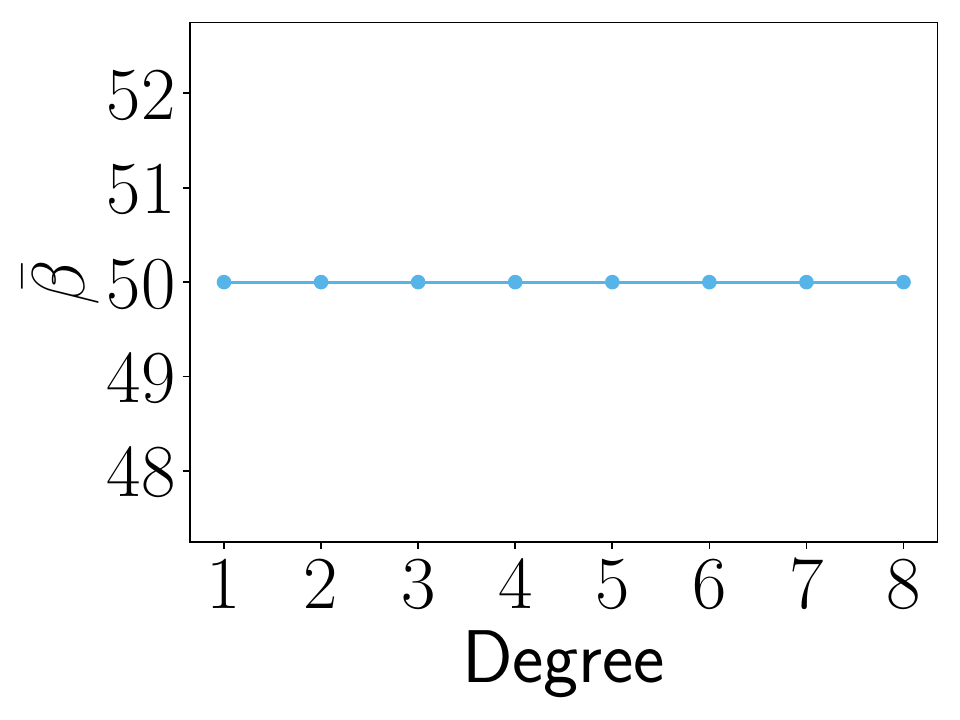}
    \end{subfigure}
    \begin{subfigure}{0.32\textwidth}
    \centering
    \caption{}
    \label{fig:surviv_deg}
         \includegraphics[width = 0.95\textwidth]{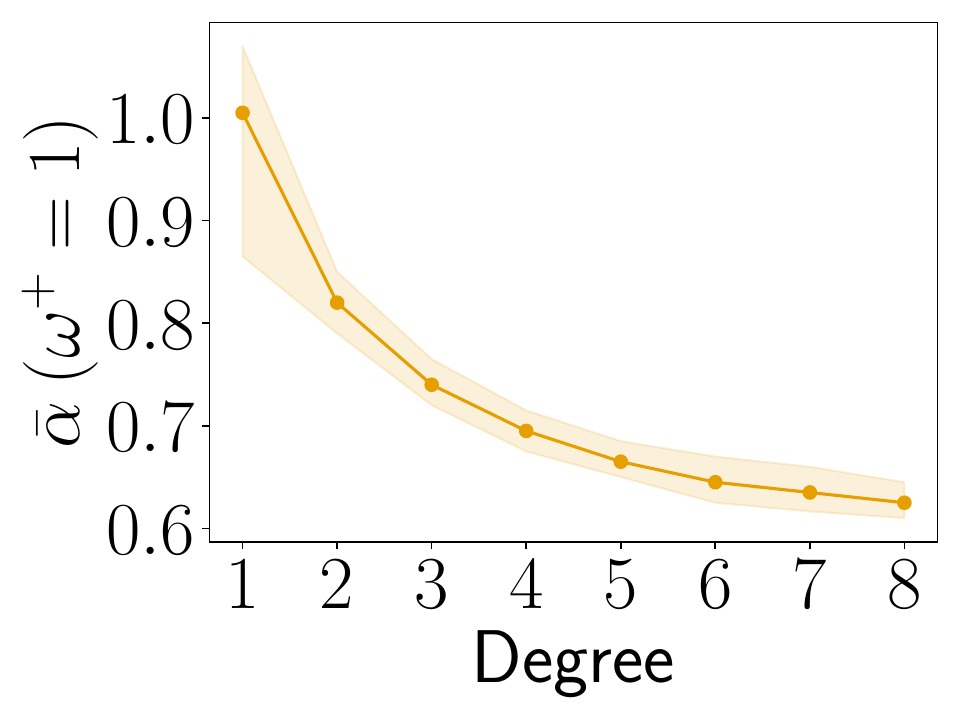}
    \end{subfigure}
    \begin{subfigure}{0.32\textwidth}
    \centering
    \caption{}
    \label{fig:surviv_deg2}
    \includegraphics[width = 0.95\textwidth]{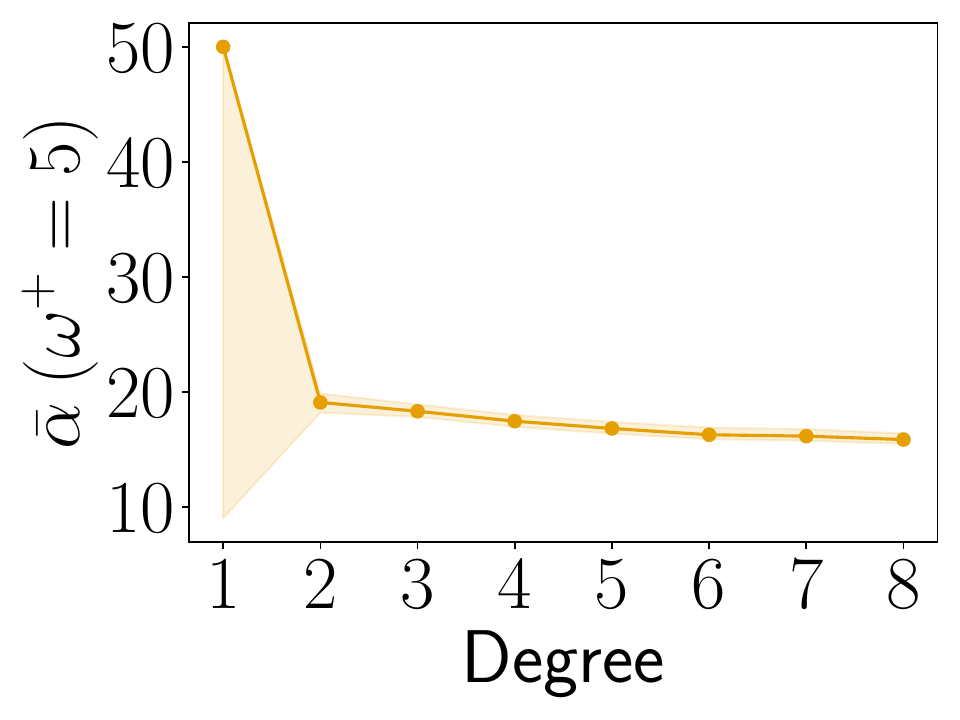}
    \end{subfigure}
    \caption{The dependence of single-node stability on degree. (a) Basin stability bound versus degree. (b) Survivability bound ($w^+=1$) versus degree. (c) Survivability bound ($w^+=5$) versus degree. The dots show the average value, and the shaded region indicates the $15.9\%$ to the $84.1\%$ percentile.}
    \label{fig:deg}
\end{figure}

Fig. \ref{fig:deg} shows the relationship between degree and single-node stability.  Fig. \ref{fig:stab_deg} does not show a correlation between basin stability bound and a node's degree. Nodes with low basin stability bound are not captured by this plot as their values lie below the $15.9\%$ percentile range;  these nodes are better examined by looking at the basin stability bound values for different network motifs. Fig. \ref{fig:surviv_deg} shows the dependence of survivability bound $( \omega^+ = 1)$ on degree --- a strong negative correlation is seen. {For small perturbations, it is numerically observed that, after the initial phase and frequency disturbances in the systems, oscillations quickly die out. Thus, for a smaller desirable region, as in the case of $\omega^+ = 1$, where perturbations inside the survivability bound are small, the initial frequency fluctuations in the system are responsible for pushing the system out of the region of survival. Consider a single-node perturbation given by $x_i^p = (\theta_1^s,\: \theta_2^s,\: ...,\: \theta_i^p,\: ..., \:\theta_N^s,\: 0, \:0, \:..., \:0,\: ...,\: 0)$. Assuming that the time period $\Delta t$ and the frequency deviations $\Delta\omega_i$ after the perturbation are sufficiently small, then $\Delta\omega_i $ can be approximated as $
\Delta\omega_i = \dot\omega_i\bigl|_{x=x_i^p}\,\Delta t$  right after the perturbation $x_i^p$, before the effect of nonlinearities and damping of the oscillations. In time $\Delta t$, the maximum frequency deviation in the grid can be written as $
\max_i \{\Delta\omega_i\} = \max_i \{\dot\omega_i\bigl|_{x=x_i^p}\}\,\Delta t$. For small single-node perturbations, the expression $\max_i \{\dot\omega_i\bigl|_{x=x_i^p}\}$, where $\dot\omega_i$ is given by Eq. \ref{eq:eqswing_simple_b}, is positively correlated with a nodes degree, thus making the initial grid frequency deviation $\max_i \{\Delta\omega_i\}$ higher for nodes with higher degrees. Thus, when $\omega^+$ is small, smaller perturbations are required at high-degree nodes to keep frequency deviations under the bound $\omega^+$ --- this explains the trend observed in Fig. \ref{fig:surviv_deg}.}

\begin{figure}[tbh]
    \centering
    \begin{subfigure}{0.32\textwidth}
    \centering
    \caption{}
    \label{fig:stab_class}
    \includegraphics[width = 0.95\textwidth]{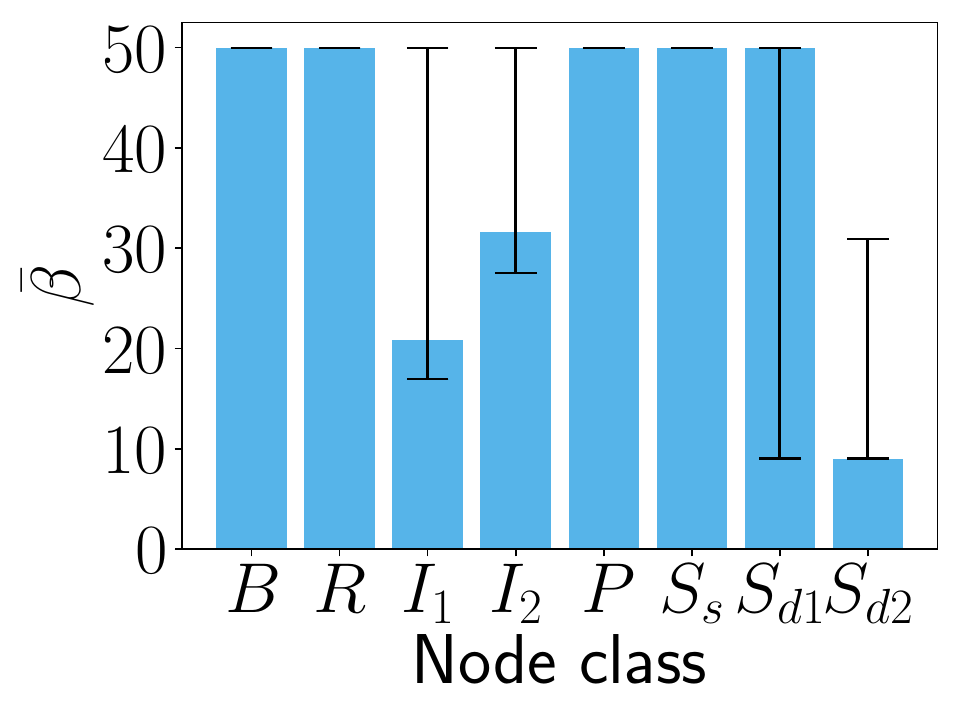}
    \end{subfigure}
    \begin{subfigure}{0.32\textwidth}
    \centering
    \caption{}
    \label{fig:surviv_class}
         \includegraphics[width = 0.95\textwidth]{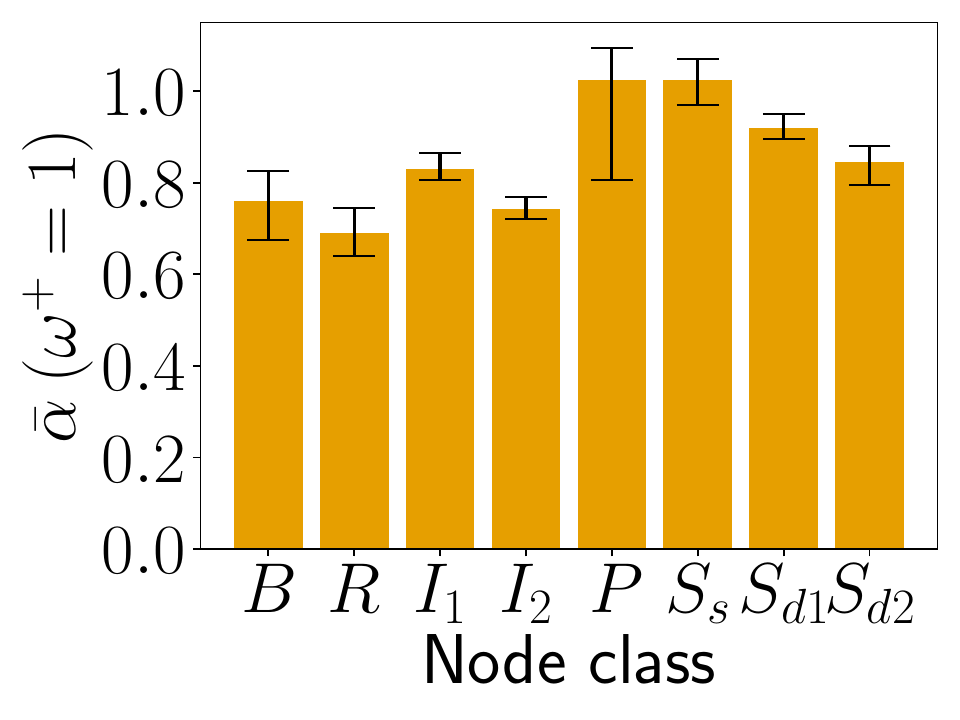}
    \end{subfigure}
    \begin{subfigure}{0.32\textwidth}
    \centering
    \caption{}
    \label{fig:surviv_class2}
         \includegraphics[width = 0.95\textwidth]{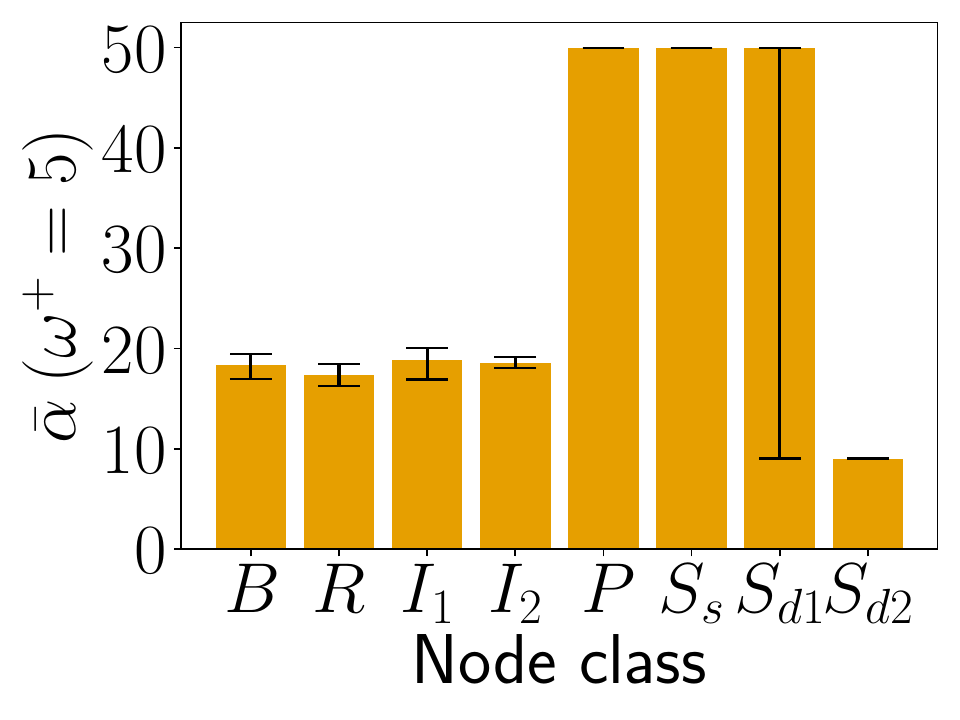}
    \end{subfigure}
    \caption{Bar graph showing the average single-node stability for {bulk nodes ($B$), root nodes ($R$), class 1 inner tree nodes ($I_1$), class 2 inner tree nodes ($I_2$), proper leaves ($P$), sparse sprouts ($S_s)$, class 1 dense sprouts ($S_{d1}$) and class 2 dense sprouts ($S_{d2}$).} (a) Basin stability bound for each node class. (b) Survivability bound ($w^+=1$) for each node class. (c) Survivability bound ($w^+=5$) for each node class. The black line indicates the $15.9\%$ to the $84.1\%$ percentile. }
    \label{fig:class}
\end{figure}

Fig. \ref{fig:surviv_deg2} also shows a negative correlation with survivability bound $( \omega^+ = 5)$ and degree. However, nodes with a degree of one take a wide range of survivability bound values. Nodes of degree one can be further bifurcated into proper leaves, sparse sprouts, class 1 dense sprouts, and class 2 dense sprouts. These node classes are examined in Fig. \ref{fig:class}. 

Fig. \ref{fig:class} shows the average single-node stability values for the different classes of nodes described in section \ref{sec:mot}. In Fig. \ref{fig:stab_class}, we observe that class 1 and class 2 dense sprouts have the lowest basin stability bound values out of all the nodes with a degree of one. Additionally, Fig. \ref{fig:stab_class} shows that class 1 and class 2 inner tree nodes have low basin stability bound values. In Fig. \ref{fig:surviv_class}, we observe that root nodes and bulk nodes have the lowest survivability bound values (with $w^+=1$). These nodes are in the interior of networks and thus have a higher degree than other nodes. This agrees with the fact that survivability bound is negatively correlated with a node's degree (Fig. \ref{fig:surviv_deg}). In Fig \ref{fig:surviv_class2}, it is observed that dense sprouts have the lowest survivability bound values (with $w^+=5$). Thus, as the allowed frequency deviations increase, dense sprouts move from having relatively high survivability values compared to other nodes to having the lowest survivability values. 

{The plots in Fig. \ref{fig:class} capture the average single-node stability for different node classes; however, these plots do not account for outlier values that lie below the $15.9 \%$ percentile. To investigate the most vulnerable nodes in a network, examining the stability of outlier nodes is required. To examine the stability of the least stable nodes for each node class, we define the following metric known as tail expectation:
\begin{equation}
    T_{c, \:p}  = \mathbb{E}[X_c | X_c \leq q_{c, \: p}] 
\end{equation}
Here, $X_c$ is a random variable that takes on stability values that follow the distribution of the computed stability values of the node class $c$, and $q_{c, \: p}$ is the value of the $p^\text{th}$ percentile of the distribution. For the analysis, we choose $p = 3 \%$, implying $T_{c, \: p}$ is the expected value of stability for the lowest $3\%$ of stability values.}

\begin{figure}[tbh]
    \centering
    \begin{subfigure}{0.4\textwidth}
    \centering
    \caption{}
    \label{fig:stab_class_tail}
    \includegraphics[width = 0.95\textwidth]{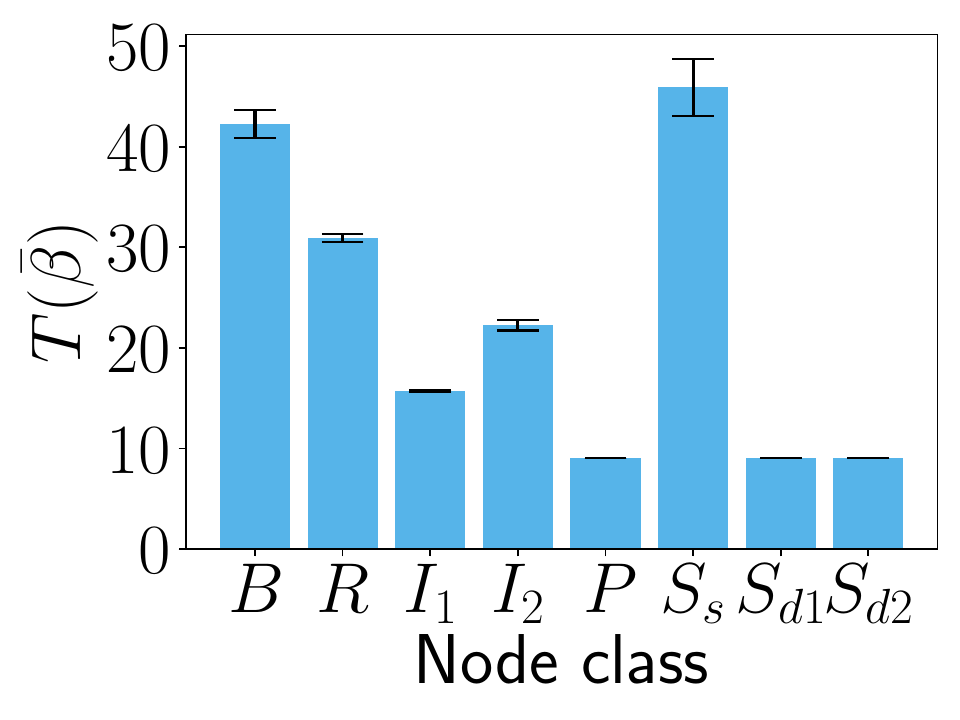}
    \end{subfigure}
    \begin{subfigure}{0.4\textwidth}
    \centering
    \caption{}
    \label{fig:surviv_class_tail}
         \includegraphics[width = 0.95\textwidth]{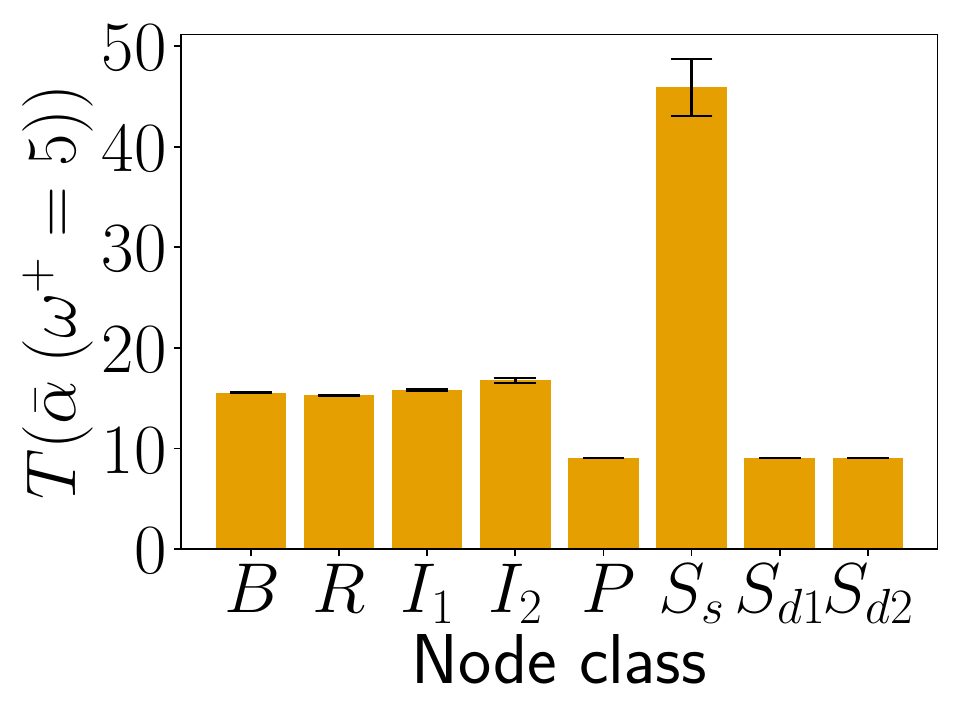}
    \end{subfigure}

    \caption{Bar graph showing the tail expectation of single-node stability for {bulk nodes ($B$), root nodes ($R$), class 1 inner tree nodes ($I_1$), class 2 inner tree nodes ($I_2$), proper leaves ($P$), sparse sprouts ($S_s)$, class 1 dense sprouts ($S_{d1}$) and class 2 dense sprouts ($S_{d2}$).} (a) Basin stability bound for each node class. (b) Survivability bound ($w^+=5$) for each node class. The black line indicates the confidence interval corresponding to one standard error.} 
    \label{fig:class2}
\end{figure}

{Using this metric, we obtain plots (Fig. \ref{fig:class2}) for basin stability bound and survivability bound $(\omega^+ = 5)$.}
{In Fig. \ref{fig:stab_class_tail} and Fig. \ref{fig:surviv_class_tail}, the tail expectation of both basin stability bound and survivability bound is the lowest for proper leaves and dense sprouts, indicating that the nodes with the lowest stability are proper leaves and dense sprouts. 

In Fig. \ref{fig:traj}, we have perturbed the most unstable proper leave nodes and dense sprouts with perturbations of strengths equal to their tail expectations. It is observed that the perturbation to the proper leave node leaves the proper leave oscillating with a distinctively different frequency, while the other nodes in the network oscillate closely around the grid frequency. This implies that, after a perturbation, control to the proper leave alone could lead to the grid being resynchronized. On the other hand, the perturbation to the dense sprout causes multiple nodes to oscillate at different frequencies leaving the grid in a desynchronized state described by J Nitzbon et al. \cite{nitzbon2017deciphering}. 

\begin{figure}[h]
    \centering
    \begin{subfigure}{0.45\textwidth}
    \centering
    \caption{}
    \label{fig:traj1}
    \includegraphics[width = 0.95\textwidth]{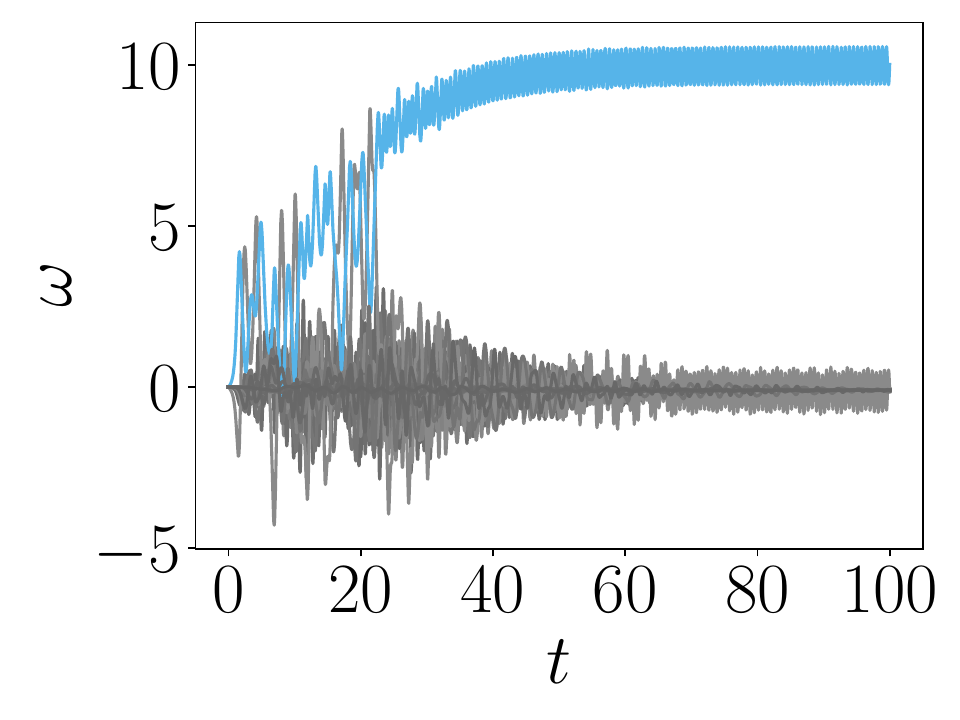}
    \end{subfigure}
    \begin{subfigure}{0.45\textwidth}
    \centering
    \caption{}
    \label{fig:traj2}
         \includegraphics[width = 0.95\textwidth]{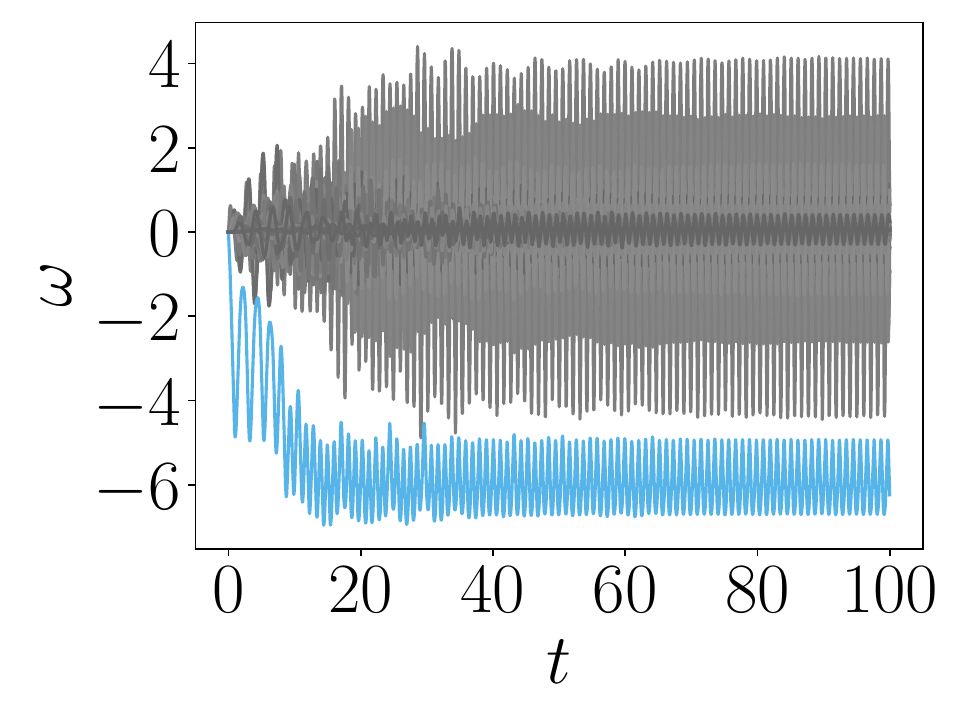}
    \end{subfigure}
    
    \caption{{Plot showing the frequency ($\omega$) versus time ($t$) of all nodes in a network for a perturbation to a (a) proper leave and a (b) dense sprout. The blue curve is the trajectory of the perturbated node, while the grey curves are the trajectories of the other nodes in the network.}} 
    \label{fig:traj}
\end{figure}

Previous studies that have examined power grid stability using a Kuramoto-like model have identified the heightened vulnerability of inner leaves (based on betweenness centrality) \cite{menck2014dead}, as well as dense sprouts \cite{nitzbon2017deciphering}. However, the vulnerability of proper leaves has not been observed before. In this study, using probabilistic distance-based stability quantifiers, we have identified proper leaves and dense sprouts to be the most vulnerable in the case of large perturbations and large allowed grid frequency deviations. }

\section{Conclusion}\label{sec:conclusion}

In this work, we have employed asymptotic and transient stability quantifiers to measure the magnitude of dangerous single-node power grid perturbations. A perturbation stronger than a node's basin stability bound can be devastating for the grid, and can result in grid synchrony loss and cascading failure. On the other hand, a perturbation stronger than a node's survivability bound might be undesirable for the grid as it momentarily pushes node frequencies out of the desirable operation region without necessarily pushing the grid permanently out of synchrony.   

We have studied the effect of network topology on single-node stability using these distance-based measures of stability, departing from commonly used volume-based stability quantifiers. Through the study of the single-node stability of an ensemble of synthetic power grids, we have identified vulnerable local network properties of power grids. We have found that {proper leaves} and dense sprouts have the lowest values of basin stability bound. Hence, large perturbations to these nodes should be avoided at all costs. On the other hand, nodes with a high degree have low survivability bound values (when the tolerated frequency deviations in the grid are small ($\pm 0.64$ Hz)). This means that perturbations to these nodes have the highest chance of resulting in undesirable transients in the grid but may rarely ever result in total loss of synchrony due to such nodes having high basin stability bound values. However, when the tolerated frequency deviations are larger (($\pm 3.2$ Hz), {proper leaves} and dense sprouts have the lowest survivability values. 

In this paper, we have used distance-based nonlinear stability methods on power grids, however, they are readily applicable to other complex-network dynamical systems. We believe that the methods used and the results obtained in this study can be useful in future research on the stability of power grids and other complex networks.

\section*{Acknowledgments}
We thank Dr. Frank Hellmann from the Potsdam Institute for Climate Impact Research for his helpful suggestions. S B acknowledges the J C Bose National Fellowship
provided by SERB, Government of India, Grant No. JBR/2020/000049.

\section*{ORCID}
\noindent Calvin Alvares - \url{https://orcid.org/0009-0009-9113-1922}

\noindent Soumitro Banerjee - \url{https://orcid.org/0000-0003-3576-0846}

%\markboth{Authors' Names}{Paper Title}

%\bibliographystyle{apalike}  % Uses author-year formatting
%\bibliography{bibliography}  % Bibliography file %(references.bib)

\end{document}